\documentclass{jetpl}
\twocolumn

\title{QCD string and the Lorentz nature of confinement}
\rtitle{QCD string and the Lorentz nature of confinement}
\author{A.\,V.\,Nefediev, Yu.\,A.\,Simonov}
\address{Institute of Theoretical and Experimental Physics, B.Cheremushkinskaya 25, 117218, Moscow, Russia}
\rauthor{A.\,V. Nefediev, Yu.\,A.\,Simonov}

\abstract{We address the question of the Lorentz nature of the effective long--range interquark interaction generated 
by the QCD string with quarks at the ends. Studying the Dyson--Schwinger equation for a heavy--light quark--antiquark system, 
we demonstrate explicitly how a Lorentz--scalar interaction appears in the Diraclike equation for the light quark,
as a consequence of chiral symmetry breaking.
We argue that the effective interquark interaction in the Hamiltonian of the QCD string with quarks at the ends 
stems from this effective scalar interaction.}


\begin{document}
\maketitle

\section{Introduction}

Description of the spectrum of mass and other properties of hadrons is one of the main tasks of QCD, as the theory of strong interactions,
and a variety of nonperturbative theoretical approaches are developed for this purpose. In this paper we touch upon two of them, which we
believe to be complementary to one another. On one hand, the quantum--mechanical approach of the QCD string with quarks at the ends can be derived,
starting from the fundamental QCD Lagrangian. On the other hand, a field--theory--inspired approach based on the Dyson--Schwinger equation for
quarkonia can be developed. The two mentioned approaches allow one to have reliable predictions for properties of hadrons, although both of
them meet certain problems and should be applied together, side by side. For example, the approach based on the Dyson--Schwinger equation for
quarkonia is
well adjusted for the case of heavy--light systems, whereas in the light--light case its application is not straightforward. In the meantime,
the QCD string approach can be readily applied to both heavy--light and light--light quarkonia, as well as to other hadrons, including those
with excited gluonic degrees of freedom. Unfortunately, since the effects of spontaneous breaking of chiral symmetry (SBCS) are not inherent to
the string model, there is no hope to reproduce the properties of the lightest states in the mesonic spectrum --- the pions and the kaons ---
in this approach. This is the problem for which the method of the Dyson--Schwinger equation comes to rescue. 
Indeed, this method appears appropriate for studies of chiral symmetry breaking (CSB) and for all phenomena related to it.
In this paper we make one more step on the way of merging the two aforementioned methods and study the problem of the Lorentz nature of
confinement generated by the QCD string. Considering a heavy--light quarkonium, we employ the Dyson--Schwinger approach to derive an effective
Diraclike bound--state equation for the spectrum of the system. We argue that it is SBCS, caused by confinement, which
gives rise to the scalar part of the interquark interaction in this equation and demonstrate how the confinement--induced scalar interaction reduces, under
certain conditions, to a local potential dynamics described by the quantum--mechanical Salpeter equation for the quarkonium. 
Finally, we extend this conclusion of the scalar nature of the effective interquark interaction to the case of the rotating QCD string 
with quarks at the ends.

\section{QCD string and the spinless Salpeter equation}

In this section we remind the reader the main steps used to derive the Hamiltonian of the QCD string with quarks at the ends in the Vacuum
Correlator Method (VCM) \cite{1}. We start from the gauge--invariant in-- and out--states of the quarkonium,
$\Psi^{({\rm in,out})}_{q\bar q}(x,y|A)=\bar{\Psi}_{\bar q}(x)P\exp{\left(ig\int_{y}^{x}dz_{\mu}A_{\mu}\right)}\Psi_q(y)$ \cite{2}.
Now, writing the Green's function of the (flavour--nonsinglet) quark--antiquark meson, 
\begin{equation}
G_{q\bar q}=\langle\Psi_{q\bar q}^{(\rm out)}(\bar{x},\bar{y}|A){\Psi^{(\rm in)\dagger}_{q\bar q}}(x,y|A)\rangle_{q\bar{q}A},
\end{equation}
and performing averaging over the gluonic field, by means of the minimal area law for the
isolated Wilson loop, we can extract the standard Nambu--Goto effective action for the string connecting the quarks,
\begin{multline}
S_{\rm min}=\int_0^Tdt\int_0^1d\gamma\sqrt{(\dot{w}w')^2-\dot{w}^2{w'}^2},\\
w_{\mu}(t,\gamma)=\gamma x_{1\mu}(t)+(1-\gamma)x_{2\mu}(t),
\label{str0}
\end{multline}
where we used the straight--line string ansatz for the minimal surface \cite{2}. Finally, considering the quark--antiquark system at rest in the
laboratory reference frame, we synchronise the quark times and put them equal to the laboratory time, $x_{10}=x_{20}=t$. The resulting centre--of--mass
Hamiltonian reads \cite{2}:
\begin{multline}
H=\sum_{i=1}^2\left[\frac{p_r^2+m_i^2}
{2\mu_i}+\frac{\mu_i}{2}\right]+\int^1_0d\gamma\left[\frac{\sigma^2r^2}{2\nu}+
\frac{\nu}{2}\right]\\
+\frac{\vec{L}^2}{2r^2[\mu_1{(1-\zeta)}^2+\mu_2{\zeta}^2+
\int^1_0d\gamma\nu{(\gamma-\zeta)}^2]},\\
\zeta=\frac{\mu_1+\int^1_0d\gamma\nu\gamma}{\mu_1+\mu_2+\int^1_0d\gamma\nu},
\label{Hm}
\end{multline}
where, in order to get rid of the square roots in the relativistic quark kinetic terms and in the string term (\ref{str0}), we used the auxiliary field method and
introduced the einbeins $\mu_{1,2}$ and $\nu(\gamma)$. The interested reader can find the details of the einbein field formalism in the
original paper \cite{4} and examples of its application to the QCD string with quarks at the ends in Refs.~\cite{2,5}. 
Notice that extremum conditions for all
three einbeins are understood in order to arrive at the Hamiltonian in the final form, expressed via physical degrees of freedom only.
The spinless Hamiltonian
(\ref{Hm}) is to be supplied by the nonperturbative spin--orbit interaction \cite{6} as well as by the Coulomb potential and spin--dependent
terms generated by the latter. The resulting model appears rather successful in studies of quarkonia --- for example, the spectrum of
heavy--light $D$, $D_s$, $B$, and $B_s$ mesons can be reproduced with a good accuracy this way \cite{7}.
The last, angular--momentum--dependent, term in the Hamiltonian (\ref{Hm}) contains a strong contribution of the proper dynamics of the
string, described by the integral term in the denominator. The effect of this dynamics over the properties of the system 
is comprehensively studied in the literature and is known to bring the Regge trajectories slope to the
experimental value \cite{2,3}, to lower the masses of orbitally excited states \cite{7}, and so on. 
In the meantime, this contribution does not affect the Lorentz nature of the interquark interaction generated by the string.
Therefore, for the sake of simplicity, we restrict ourselves with the case of $L=0$ in the Hamiltonian (\ref{Hm}) and then 
we take extrema in all einbeins explicitly. The resulting Hamiltonian, 
\begin{equation}
H=\sqrt{p_r^2+m_1^2}+\sqrt{p_r^2+m_2^2}+\sigma r,
\end{equation}
gives rise to the well--known Salpeter equation for the spectrum. For the heavy--light system with $m_1\equiv M\to\infty$ and $m_2\equiv m$ 
it reads:
\begin{equation}
[\sqrt{p_r^2+m^2}+\sigma r]\psi(r)=E\psi(r),
\label{Salp10}
\end{equation}
where $E$ describes the excess of the bound--state energy over the heavy--quark mass.
Eq.~(\ref{Salp10}) is usually referred to as the Salpeter equation with the Lorentz--vector interaction \cite{12}, as opposed
to the would-be Lorentz--scalar confinement, as in the equation
\begin{equation}
[\sqrt{p_r^2+(m+\sigma r)^2}]\psi(r)=E\psi(r).
\label{Salp12}
\end{equation}
Therefore, according to general expectations, the Klein paradox might have operated for such a system,
and one might have expected
problems with the collapse of the mesonic wave functions and uncontrolled production of light--quark pairs by such an interaction, if
confinement had been present in the effective Dirac equation for the light quark in the form of a Lorentz time vector.
The aim of the present paper is to argue that this conclusion is misleading in the sense that the form (\ref{Salp10}) of the Salpeter
equation {\em does not imply} that the confining potential $\sigma r$ appears as a Lorentz--vector interaction in the one--particle
Dirac equation for the light quark. On the contrary, we demonstrate that an effective scalar interquark interaction
appears in this equation as a result of CSB, nevertheless the resulting Salpeter equation having the form of Eq.~(\ref{Salp10}), rather
than of Eq.~(\ref{Salp12}). As far as Eq.~(\ref{Salp12}) is concerned, it was demonstrated in Ref.~\cite{12} that its spectrum contradicts
the phenomenology of heavy--light mesons. Notice also that we are not aware of any consistent way to derive such an equation in QCD.

\section{Heavy--light quarkonium in the Dyson--Schwinger approach}

We start in this chapter with the necessary details of the Dyson--Schwinger approach to heavy--light quarkonium suggested in Ref.~\cite{13}.
Since the trajectory of the infinitely heavy particle is straight--line, then it is convenient to fix the so-called modified Fock--Schwinger
gauge \cite{14} for the background gluonic field (we work in Euclidean space),
\begin{equation}
\vec{x}{\vec A}(x_4,\vec{x})=0,\quad A_4(x_4,{\vec 0})=0,
\label{gauge}
\end{equation}
and thus to reduce the role of the static antiquark to providing the overall gauge invariance of the $q\bar q$ Green's function which, in the
gauge (\ref{gauge}), coincides with the Green's function of the light quark. Then the Dyson--Schwinger equation can be derived for the 
latter \cite{13},
\begin{multline}
(-i\hat{\partial}_x-im)S(x,y)+\int d^4z\gamma_4S(x,z)\gamma_4{\cal K}(x,z)S(z,y)\\
=\delta^{(4)}(x-y),
\label{DS}
\end{multline}
where only the structure $\gamma_4\times\gamma_4$ is kept for the sake of simplicity, whereas the interaction with the full structure 
$\gamma_\mu\times\gamma_\nu$ can be studied as well (see Refs.~\cite{13,15}). The quark kernel ${\cal K}(x,y)$
is related to the profile function $D(\tau,\lambda)$,
\begin{multline}
{\cal K}(x,y)={\cal K}(x_4-y_4,\vec{x},\vec{y})\\
=(\vec{x}\vec{y})\int_0^1d\alpha\int_0^1 d\beta D(x_4-y_4,|\alpha\vec{x}-\beta\vec{y}|),
\label{kern1}
\end{multline}
which, in turn, parametrises the bilocal correlator of the gluonic--field tensors, 
$\langle F(x)F(y)\rangle\propto D(x-y)$ \cite{1}. 
The profile $D$ decreases in all directions of the Euclidean space--time with the correlation length $T_g$, for which lattice
simulations give as small value as $T_g\simeq 0.2\div 0.3 fm$ \cite{17} and, therefore, the limit $T_g\to 0$ --- known as the 
string limit of QCD --- is adequate. In this limit, the profile function $D(\tau,\lambda)$ can be approximated by the
delta--functional form, $D(\tau,\lambda)=2\sigma\delta(\tau)\delta(\lambda)$, which is consistent with the definition of the string tension
\cite{1},
\begin{equation}
\sigma=2\int_0^\infty d\tau\int_0^\infty d\lambda D(\tau,\lambda).
\end{equation}
Then, with the help of Eq.~(\ref{kern1}), the kernel is found in the following form:
\begin{multline}
K(\vec{x},\vec{y})\equiv\frac12\int_{-\infty}^\infty{\cal K}(x_4-y_4,\vec{x},\vec{y})
e^{i\omega(x_4-y_4)}d(x_4-y_4)\\
=\frac12(\vec{x}\vec{y})\int_0^1d\alpha\int_0^1 d\beta \int_{-\infty}^{\infty} d\tau D(\tau,|\alpha\vec{x}-\beta\vec{y}|)\\
\approx\frac12\sigma(|\vec{x}|+|\vec{y}|-|\vec{x}-\vec{y}|),
\label{kern2}
\end{multline}
where, in the last, approximate, equality, the requirement of strict collinearity of the vectors $\vec{x}$ and $\vec{y}$ is relaxed, which
is admissible at large distances, $|\vec{x}|,|\vec{y}|\gg |\vec{x}-\vec{y}|$. 

The ultimate form of Eq.~(\ref{kern2})
allows one to establish a link to potential quark models for QCD \cite{18} as well as to generalise the shape of the confining 
interquark interaction from linear confinement $\sigma r$ to a generic form $V(r)$.
Now we can rewrite the Dyson--Schwinger Eq.~(\ref{DS}), in Minkowski space, in the form:
\begin{equation}
({\vec \alpha}\vec{p}+\beta m)\Psi(\vec{x})+\int d^3z\Lambda(\vec{x},\vec{z})K(\vec{x},\vec{z})\Psi(\vec{z})=E\Psi(\vec{x}),
\label{DS2}
\end{equation}
where the quantity $\Lambda(\vec{x},\vec{z})$, introduced in Ref.~\cite{13}, is defined as
\begin{multline}
\Lambda(\vec{x},\vec{z})\equiv 2i\int\frac{d\omega}{2\pi}S(\omega,\vec{x},\vec{z})\beta\\
=\sum_{n=-\infty}^\infty\Psi_n(\vec{x}){\rm sign}(n)\Psi_n^\dagger(\vec{z}).
\label{Lambda}
\end{multline}
It is clear that the Lorentz nature of confinement in Eq.~(\ref{DS2}) depends entirely on the matrix structure of $\Lambda(\vec{x},\vec{z})$. 
To proceed we stick to the formalism of the chiral angle $\varphi_p$ --- the standard approach used in potential quark models \cite{18}.
In this formalism, the positive-- and negative--energy solutions to the bound--state Eq.~(\ref{DS2}) can be
parametrised in the form \cite{185}:
\begin{multline}
\Psi_{n>0}(\vec{p})=T_p{\psi(\vec{p})\choose 0},\quad\Psi_{n<0}(\vec{p})=T_p{0\choose \psi(\vec{p})},\\
T_p=\exp{\left[-\frac12(\vec{\gamma}\hat{\vec{p}})\left(\frac{\pi}{2}-\varphi_p\right)\right]}.
\label{def0}
\end{multline}
The wave function $\psi(\vec{p})$ obeys a Schr{\" o}dingerlike eigenvalue equation which follows from Eq.~(\ref{DS2}) after
the exact Foldy--Wouthuysen transformation generated by the Foldy operator $T_p^\dagger$ (see Eq.~(\ref{FW4}) below). 
The chiral angle $\varphi_p$ is the solution to the mass--gap equation,
\begin{multline}
\label{lmge}
p\sin\varphi_p-m\cos\varphi_p\\
=\frac{\sigma}{p^2}\int_0^{\infty}\frac{dk}{2\pi}\left[\frac{4p^2k^2}{(p^2-k^2)^2}\sin[\varphi_k-\varphi_p]
\left(\frac{2pk}{(p+k)^2}\right.\right.\\
+\left.\left.\ln\left|\frac{p-k}{p+k}\right|\right)\cos\varphi_k\sin\varphi_p\right],
\end{multline}
quoted here without derivation for the linearly rising potential.
The interested reader can find the details of this formalism in Ref.~\cite{18}. 
Notice that the chiral angle also plays the role of the Foldy angle, and this is a general feature of such models.
For the purpose of the present research it is sufficient to bear in mind that the chiral angle is a continuous
smooth function which starts from
$\frac{\pi}{2}$ at the origin, with the slope inversely proportional to the scale of the CSB generated by this 
solution. In the large--momentum limit, $\varphi_p$ approaches zero. It is an easy task now to compute the function $\Lambda$ \cite{185}:
\begin{multline}
\Lambda(\vec{p},\vec{q})=(2\pi)^3\delta^{(3)}(\vec{p}-\vec{q})U_p,\\
U_p=T_p^2\beta=\beta\sin\varphi_p+({\vec \alpha}\hat{\vec{p}})\cos\varphi_p,
\label{L2}
\end{multline}
and to rewrite Eq.~(\ref{DS2}) in the form:
\begin{multline}
E_pU_p\Psi(\vec{p})+\frac12\int\frac{d^3k}{(2\pi)^3}V(\vec{p}-\vec{k})\\
\times(U_p+U_k)\Psi(\vec{k})=E\Psi(\vec{p}),
\label{Se10}
\end{multline}
where $E_p$ stands for the quark dispersive law and, for the linearly rising potential, $V(\vec{p})=-\frac{8\pi\sigma}{p^4}$.
Alternatively this equation can be arrived at as the one--particle limit of the Bethe--Salpeter equation for the quark--antiquark meson
in the framework of the potential quark models \cite{18}. The Foldy--Wouthuysen transformation of Eq.~(\ref{Se10}), performed with 
the help of the Foldy operator $T_p^\dagger$, leads one to the Schr{\" o}dingerlike equation \cite{185},
\begin{multline}
E_p\psi(\vec{p})+\int\frac{d^3k}{(2\pi)^3}V(\vec{p}-\vec{k})\left[C_pC_k\right.\\
+\left.({\vec \sigma}\hat{\vec{p}})({\vec \sigma}\hat{\vec{k}})S_pS_k\right]\psi(\vec{k})=E\psi(\vec{p}),
\label{FW4}
\end{multline}
where $C_p=\cos\frac12(\frac{\pi}{2}-\varphi_p)$ and $S_p=\sin\frac12(\frac{\pi}{2}-\varphi_p)$;
${\vec \sigma}$ are Pauli matrices, and $\hat{\vec{p}}$ and $\hat{\vec{k}}$ are the unity vectors for $\vec{p}$ and $\vec{k}$, respectively.

With Eq.~(\ref{Se10}) in hands we are in the position to comment on the Lorentz nature of confinement.
CSB means an existence of the quark Green's function having the effective mass operator with the matrix $\gamma_0$ to an even power or, 
in the language of the chiral angle, the existence of a nontrivial solution to the mass--gap equation (\ref{lmge}).
This is trivially achieved for heavy quarks, when the chiral symmetry is broken explicitly, since the quark mass term provides the required 
behaviour of the quark Green's function and, in the meantime, 
saturates the chiral angle. For light (massless) quarks, such solutions for the quark Green's function and for the chiral angle are to
appear selfconsistently in order to provide SBCS. In either case the chiral angle is different from zero
for $0\leqslant p\lesssim\Lambda_\chi$, with the CSB scale $\Lambda_\chi$ given by the quark mass, for heavy quarks, and
by the nonperturbative scale $\sqrt{\sigma}$, for light quarks. The full structure of the matrix $\Lambda$
and its influence on highly excited states in the spectrum is studied in detail in Ref.~\cite{185}. For the
purpose of the present qulitative research it is sufficient to stick either to very heavy quarks or to extremely strong confinement 
(the so-called pointlike limit of $\sqrt{\sigma}\to\infty$, which can also be called the \lq\lq heavy"--string
limit). In this case, the chiral angle is $\varphi_p=\frac{\pi}{2}$ for all $p$'s, so that $U_p\approx\beta$ and, therefore,
\begin{equation}
\Lambda(\vec{x},\vec{z})\approx\beta\delta^{(3)}(\vec{x}-\vec{z}).
\label{L3}
\end{equation}
Thus, using Eqs.~(\ref{kern2}), (\ref{DS2}), and (\ref{L3}) altogether, we arrive at the effective 
Dirac equation for the light quark with a purely scalar confinement,
\begin{equation}
[{\vec \alpha}\vec{p}+\beta(m+V(r))]\Psi(\vec{x})=E\Psi(\vec{x}).
\label{De4}
\end{equation}
This coincides with the findings of Ref.~\cite{13}, where a summation of quasiclassical eigenvalues of 
Eq.~(\ref{De4}), with $V(r)=\sigma r$, was performed explicitly and the relation (\ref{L3}) was derived for light quarks.
Notice that for massless quarks and had SBCS not have happened, the chiral angle would have been identically zero, 
and the term proportional to the matrix $\beta$ in $U_p$, as it follows from Eq.~(\ref{L2}), would have vanished.
We see therefore that the effective scalar interquark interaction arises due to CSB, both explicit or
spontaneous. In the same limit of $\varphi_p=\frac{\pi}{2}$, one has $C_p=1$ and $S_p=0$, so that the interaction part of Eq.~(\ref{FW4}) 
reduces to the potential $V(r)$, in coordinate
space. As for the kinetic term in Eq.~(\ref{FW4}), for heavy quarks, it can be well approximated by the free--quark energy,
$\sqrt{\vec{p}^2+m^2}$. 
For light quarks such a substitution is more arguable, though it is known to work rather well for heavy--light as well as for excited
light--light mesons, when the nontrivial low--momentum behaviour of the dressed--quark dispersive law $E_p$ does not play a considerable 
role (notice that this approximation fails completely for the lowest light--light quark--antiquark
state --- for the chiral pion. The latter cannot be described by the Salpeter Hamiltonian and one is to consider the full 
Dyson--Schwinger equation --- see Refs.~\cite{18,181} for two complementary approaches to the problem of the pion).
Thus, starting from Eq.~(\ref{FW4}), we arrive at the Salpeter equation, 
\begin{equation}
[\sqrt{\vec{p}^2+m^2}+V(r)]\psi(\vec{x})=E\psi(\vec{x}),
\end{equation}
which, for $L=0$ and $V(r)=\sigma r$, coincides with Eq.~(\ref{Salp10}).

It is clear from Eq.~(\ref{FW4}) that the interaction term is always added to the
entire kinetic energy of the quark,
so that the resulting Salpeter equation in the form of Eq.~(\ref{Salp10}), rather than in the form of Eq.~(\ref{Salp12}), 
should not come as a surprise. Moreover, for massless quarks and no CSB, $\varphi_p=0$ everywhere, so that $C_p=S_p=\frac{1}{\sqrt{2}}$
and the interaction in Eq.~(\ref{FW4}) acquires a rather complex structure which does not reduce to a
plain potential and supports parity doublers.  
Indeed, in the resulting equation, eigenstates with opposite parity, given by $\psi(\vec{p})$ and $(\vec{\sigma}\hat{\vec{p}})\psi(\vec{p})$, come in
pairs degenerate in mass \cite{185} --- the feature inherent to vectorial interaction.
In any case, the interaction given by the matrix (\ref{L2}) does not contain Lorentz time--vector part, 
which could have been dangerous from the point of view of the Klein paradox.

\section{Conclusions}

In this paper, we, using the Dyson--Schwinger 
approach to heavy--light quarkonia, derive the effective one--particle equation for the light 
quark in the field of the static antiquark, Eq.~(\ref{DS2}). In the heavy--quark limit of $m\to\infty$ or in the \lq\lq
heavy"--string limit of $\sqrt{\sigma}\to\infty$, the chiral angle is $\varphi_p=\frac{\pi}{2}$ and
this equation reduces to the Dirac equation with purely 
scalar confinement. In the meantime, if the exact Foldy--Wouthuysen transformation is performed over the
bound--state Eq.~(\ref{DS2}), the Schr{\" o}dingerlike Eq.~(\ref{FW4}) arises with the interaction having a
rather complex structure. In the same limiting case of $\varphi_p=\frac{\pi}{2}$, this interaction can be
considerably simplified, reducing to a local potential. The spinless 
Salpeter Eq.~(\ref{Salp10}) is derived this way, with the interaction term added to the entire 
quark kinetic energy. We notice also that, without CSB, explicit or spontaneous, the chiral angle would
vanish identically giving rise to a nonpotential dynamics (as it follows from Eq.~(\ref{FW4}) with $\varphi_p=0$)
which has nothing to do with the dynamics described by the Salpeter Eq.~(\ref{Salp10}) and the like.
As the approach of the QCD string with quarks at the ends is the generalisation of the
simple potential Salpeter equation to the case of the interquark interaction incorporating the proper dynamics
of the string, then we conclude that the effective interquark interaction generated by the QCD
string has scalar nature, that is, it appears entirely due to CSB. The detailed analysis of the Lorentz nature
of confinement in quarkonia with the actual form of the chiral angle --- solution to the mass--gap equation
--- lies beyond the scope of the present paper but we emphasize that, in order to have an accurate and 
selfconsistent approach, one is to consider and 
solve the full Dyson--Schwinger Eq.~(\ref{DS}). Notice that, although the fundamental colour interaction in QCD mediated by gluons is manifestly
vectorial, this does not automatically give rise to the Salpeter Eq.~(\ref{Salp10}). 
Indeed, the effective interquark interaction, which appears after integrating out gluonic degrees of freedom and which can be  
described naturally, for example, with the help of the Diraclike Eq.~(\ref{DS2}), appears dynamically and thus no {\em a priori}
conclusion can be made concerning its Lorentz nature. We argue, therefore, that one should be careful using the notions 
of \lq\lq vector" and \lq\lq scalar" 
confinement, always giving an explicit reference to the corresponding Diraclike equation or Hamiltonian. 

The authors would like to thank Yu. S. Kalashniko\-va, M. G. Olsson, and E. Ribeiro for reading the manuscript and critical comments. 
The financial support of the grant NS-1774.2003.2, as well as of the Federal Programme of the Russian Ministry of Industry, Science, and
Technology No 40.052.1.1.1112 is acknowledged. Work of A. V. N. is also supported through the
DFG 436 RUS 113/820/0-1 and RFFI 05-02-04012-NNIOa grants.

\end{document}